# Multi-axis fields boost SABRE hyperpolarization via new strategies

Jacob R. Lindale[1], Loren L. Smith[1*], Mathew W. Mammen[2*], Shannon L. Eriksson[1,3], Lucas Everhart[1], Warren S. Warren[4]

[1] Department of Chemistry, Duke University, Durham, NC.
[2] Department of Physics, Duke University, Durham, NC.
[3] School of Medicine, Duke University, Durham, NC.
[4] Departments of Chemistry, Physics, Biomedical Engineering, and Radiology, Duke University, Durham, NC.

**The inherently low signal-to-noise ratio of NMR and MRI is now being addressed by hyperpolarization methods. For example, iridium-based catalysts that reversibly bind both parahydrogen and ligands in solution can hyperpolarize protons (SABRE) or heteronuclei (X-SABRE) on a wide variety of ligands, using a complex interplay of spin dynamics and chemical exchange processes, with common signal enhancements between $10^3$-$10^4$. This does not approach obvious theoretical limits, and further enhancement would be valuable in many applications (such as imaging mM concentration species in vivo). Most SABRE/X-SABRE implementations require far lower fields (μT-mT) than standard magnetic resonance (>1T), and this gives an additional degree of freedom: the ability to fully modulate fields in three dimensions. However, this has been underexplored because the standard simplifying theoretical assumptions in magnetic resonance need to be revisited. Here we take a different approach, an evolutionary strategy algorithm for numerical optimization, Multi-Axis Computer-aided HEteronuclear Transfer Enhancement for SABRE (MACHETE-SABRE). We find nonintuitive but highly efficient multi-axial pulse sequences which experimentally can produce a 10-fold improvement in polarization over continuous excitation. This approach optimizes polarization differently than traditional methods, thus gaining extra efficiency.**

Nuclear Magnetic Resonance (NMR) and its clinical counterpart Magnetic Resonance Imaging (MRI) generally suffer from inherently low signal-to-noise because the energy difference $\Delta E$ for spin flips is very small. For $^1$H at 10 Tesla, $\Delta E/k_B \approx 20\ mK$, so a room temperature sample at equilibrium typically has a fractional magnetization $< 10^{-4}$. Hyperpolarization methods circumvent this by inducing non-equilibrium spin polarization on target nuclei, generally by transfer from another source. Some of these methods were pioneered decades ago. For example, in dissolution Dynamic Nuclear Polarization (dDNP) the source of nuclear spin order is unpaired electron spins, most commonly as dopants in a low temperature glass, followed by dissolution(1). dDNP can generate ~40% spin polarization on clinically relevant targets such as $^{13}$C-pyruvate(2-4). While this remains a subject of active research, this method is expensive, relatively slow, and requires specialized equipment. Another approach, ParaHydrogen Induced Polarization (PHIP) works by incorporating parahydrogen (the singlet, nuclear antisymmetric state of the hydrogen molecule) catalytically into unsaturated molecular sites, and subsequently transfer polarization to nuclei in the target(5-7). PHIP has significantly reduced cost and equipment requirement relative to dDNP(8) but requires molecular transformation and specific catalyst dynamics, and thus is restricted to a very limited range of molecular motifs.

Here we focus on a newer method, Signal Amplification By Reversible Exchange (SABRE), a non-hydrogenative PHIP variant that utilizes reversible interactions between the target and parahydrogen mediated by an organometallic catalyst(9-12). While the original SABRE experiments targeted $^1$H hyperpolarization(9), the X-SABRE variants that followed(13-21) broadened the scope of polarization sites to heteronuclei. Of the X-SABRE methods, SABRE-SHEATH(15) (SABRE in SHield Enables Alignment Transfer to Heteronuclei) is by far the most popular, as it generates ~1-4% spin polarization on most heteronuclear targets in under a minute and is insensitive to changes in the substrate. The original implementation of SABRE-SHEATH utilized a static, microTesla magnetic field (conventionally denoted as $B_z$), but recent advances have shown that shaping $B_z(t)$ using elementary waveforms (pulses, square waves, etc.) boosts hyperpolarization by approximately 3-fold, a change that only requires use of an arbitrary waveform generator to drive the solenoid used to control the magnetic field(22, 23). Essentially, this approach matches the spin dynamics to the

**Significance**

**Magnetic resonance techniques are often limited by intrinsically low signals. Hyperpolarization methods such as Signal Amplification By Reversible Exchange (SABRE) circumvent these limitations by artificially generating higher nuclear spin polarizations. Here, we show that the SABRE signal may be boosted up to 10 times over conventional SABRE methods using shaped multi-axis magnetic fields generated by a computer-aided pulse sequence design algorithm, and explain the strategy the algorithm discovered, which differs from previous approaches.**

lifetime of the organometallic complex, thus optimizing the amount of polarization created.

Perhaps the greatest technical advantage of working at ultralow magnetic fields is that it is trivial to produce highly homogeneous magnetic fields which can be readily changed in any direction. Here we explore readily achievable multi-axis fields, and show that they can generate substantial improvements in achievable polarization past what has been demonstrated with z-axis modulation alone(22, 23). We demonstrate two different approaches. In the first case, we use circularly polarized fields with frequency near the Larmor frequency of a constant z field, and show that this produces polarization enhancements using a novel strategy to prevent back-pumping. We then extend the ideas of shaping the magnetic field and design the Multi-Axis Computer-aided HEteronuclear Transfer Enhancement (MACHETE) for SABRE pulse sequence, which yields up to a 10-fold improvement in heteronuclear polarization over SABRE-SHEATH (Fig. 1). We will discuss the computational methods used to design this pulse sequence, analyze the resulting optimized shape, and introduce the modifications that were made to experimentally implement the MACHETE-SABRE experiment.

## Results and discussion

We first discuss hyperpolarization in SABRE in the context of a model 3-spin system (Figure 2), where two hydride spins arise from the parahydrogen binding and the third spin is the single target nucleus (here labeled $L$). This suffices to make our arguments, although as shown in Figure 1, the 4-spin system with two target nuclei (e.g. concentrated $^{15}$N-pyridine) is also dramatically enhanced with the same waveform. This is also consistent with calculations based on all of the spins. As the structure of the Hamiltonian has been extensively discussed elsewhere, we focus here only on key points to understand our results.

The evolution of these states in an externally applied z-magnetic field is governed by the nuclear spin Hamiltonian ($\widehat{\mathcal{H}}$) shown, here in natural units ($\hbar = 1$):

$$\widehat{\mathcal{H}} = \omega_H(\hat{I}_{1z} + \hat{I}_{2z}) + \omega_L \hat{L}_z + 2\pi J_{HH} \hat{I}_1 \cdot \hat{I}_2 + 2\pi J_{HL} \hat{I}_1 \cdot \hat{L} \quad [1]$$

Only one hydrogen (here labeled "1") is scalar coupled to the ligand, as the coupling corresponding to a 90° H-Ir-L geometry is negligible. It is convenient to express the eight energy levels using the singlet-triplet basis set on the hydride spins ($|T_H^+\rangle, |T_H^0\rangle, |T_H^-\rangle, |S_H\rangle$) and the Zeeman basis on the target ligand ($|\alpha_L\rangle, |\beta_L\rangle$). Starting with pure parahydrogen, and in the absence of any spin dynamics on the complex except for exchange, the states $|S_H\alpha_L\rangle$ and $|S_H\beta_L\rangle$ are each 50% populated, the other six states are empty.

Figure 3 shows the spin evolution in a system modeled after acetonitrile, when the spins are continuously exposed to a constant z-field. Note that all of the terms in equation [1] conserve the net number of spins-up. Thus, any ligand hyperpolarization is counterbalanced by opposite sign spin angular momentum on hydride spins. In the accepted picture of SABRE/X-SABRE, ligand hyperpolarization is created by preferentially pumping a transition such as $|S_H\beta_L\rangle \rightarrow |T_H^-\alpha_L\rangle$ without equal excitation of the $|S_H\alpha_L\rangle \rightarrow |T_H^-\beta_L\rangle$. As discussed in detail elsewhere(23), when $J_{HH} \gg J_{HL}$ the optimal field for this imbalance is given by a level anticrossing (LAC) condition $\omega_H - \omega_L = -2\pi J_{HH} + \pi J_{HL}/2$ but in the usual X-SABRE experimental case the experimental optimum is quite different; for the $^{15}$N-acetonitrile system ($J_{NH} = -25.41\ Hz$, $J_{HH} = -8\ Hz$) the LAC occurs at $B = \pm 0.04\ \mu T$, an order of magnitude below the actual experimental maximum around 0.6 µT. Figure 3A shows the coherent dynamics (ignoring exchange). An average difference between ligand α and β populations (Fig. 3B, top) lead to some solution hyperpolarization after exchange (Fig. 3B, bottom) as rebinding events allow this polarization to build up (Fig. 3C), typically to 1-4% magnetization as noted earlier.

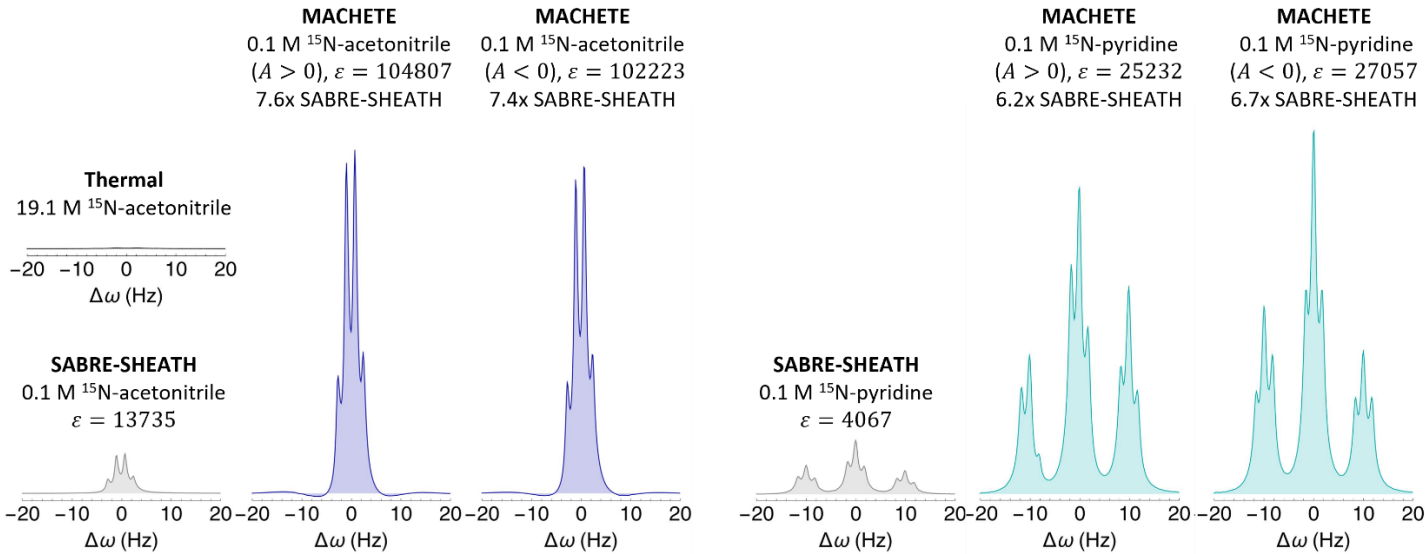

**Figure 1.** Thermal and hyperpolarized spectra of $^{15}$N-pyridine (cyan) and $^{15}$N-acetonitrile (blue). MACHETE produces an order of magnitude improvement in signal over SABRE-SHEATH ($\varepsilon_{MACHETE}/\varepsilon_{SHEATH}$) at optimal field conditions. Significant improvement over SABRE-SHEATH is observed for both the ES optimized MACHETE pulse with $A_z < 0$ and the $A_z > 0$ variant.

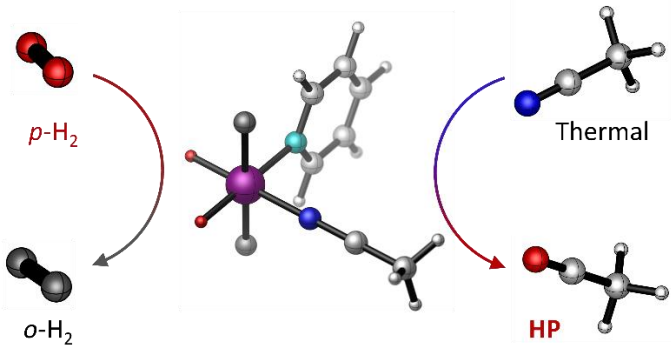

**Figure 2.** Exchange interactions in SABRE with an Ir catalyst (prototype three-spin "Y configuration" model). Parahydrogen, a target ligand (here $^{15}$N-acetonitrile), and a stabilizing coligand (here $^{14}$N-pyridine) exchange at the equatorial sites to form transient spin coupling networks for polarization transfer. In the actual catalyst, the two axial positions have ligands (commonly IMes for one position, an exchangeable ligand for the other). This is also a good model for hyperpolarization of natural abundance ligands. While we use this for illustration purposes here, our calculations generally include all spins, and the results are also applicable to the case of two bound polarizable ligands, such as concentrated $^{15}$N-pyridine, which we call the "X configuration" case.

Multiple recent papers have demonstrated methods to improve the polarization efficiency of SABRE-SHEATH experiments. Coherent SABRE uses $B_z$ fields near the optimum for a very short time (typically 15-20 ms) to access the first minima of $|S_H\alpha_L\rangle$ population in Figure 3A, then switches to a high field to slow the dynamics until exchange occurs. More complex waveforms (unbalanced square waves, sine waves, chirped fields), again only in one direction, were characterized in refs. (22, 23) using a combination of average Hamiltonian theory and numerical simulations. As ref. (23) shows, this works by effectively slowing the oscillatory dynamics (reducing the effective size of the $J_{HL}$ coupling) to improve transfer efficiency at the average exchange time, which can also be seen in the slower oscillatory dynamics.

The focus of this paper is on three-dimensional fields to improve polarization levels. It is not obvious that such an approach can work, because most of the accumulated polarization is sitting on unbound ligands in solution and the z magnetization could be destroyed by fields along other directions. However, such fields also offer the possibility (not achievable with $B_z$ fields alone) to move population out of states which would decrease population after rebinding. For example, if the $|S_H\beta_L\rangle \rightarrow |T_H^-\alpha_L\rangle$ is selectively pumped, ligand exchange followed by rebinding of a ligand spin-up will allow the $J_{HL}$ coupling to generate back-pumping to singlet, in the usual case where ligand exchange is more rapid than hydrogen exchange. However, $B_x$ or $B_y$ fields can move the population in $T_H^-$ into the other two triplet states, which do not efficiently deplete ligand magnetization.

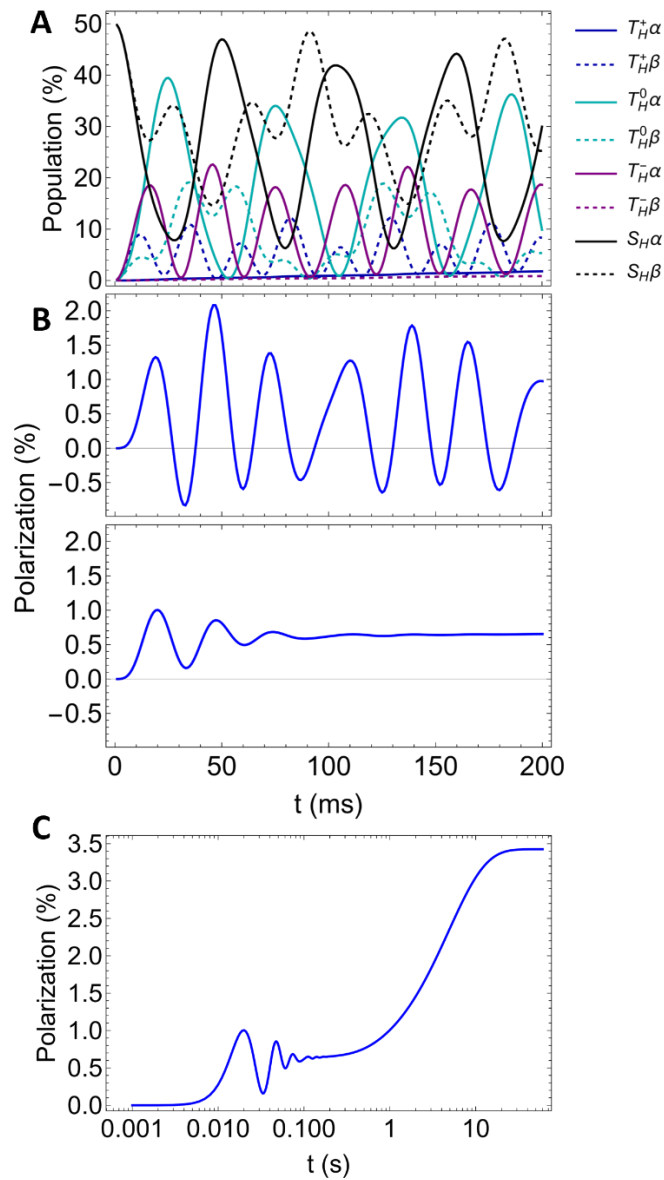

**Figure 3.** Coherent evolution (ignoring any exchange) in the SABRE-SHEATH experiment at the optimal magnetic field for polarization transfer (0.6 µT). Initially only the $S_H\alpha$ and $S_H\beta$ states are populated, but at later times the average population of the α states (dashed) is modestly higher than the β states (solid). This leads, after ligand exchange, to excess β population in solution. For the simulations with exchange, $k_L = 20\ s^{-1}$, $k_H = 1\ s^{-1}$, and $T_{1,L} = 22.5\ s$.

Recently, we introduced a physically exhaustive model for SABRE simulations(24), which has been rigorously tested against experimental systems(22, 23). This permits us to explore parameter space far faster than would be experimentally feasible and allows us to investigate the use of multi-axial fields to generate SABRE hyperpolarization. We find an efficient way to shunt population out of the triplet state targeted by the SABRE transition ($T_H^-$ for instance) and into the other triplet states, thus preventing depolarization upon exchange, was to utilize a circularly polarized transverse magnetic field with an amplitude of $B_{x,y} = B_z$ at a frequency slightly offset from $\omega = (\gamma_H + \gamma_L)B_z$, the double quantum transition. Under these conditions, simulations demonstrated significant enhancement of the SABRE hyperpolarization over SABRE-SHEATH, which we confirmed experimentally (Figure 4) using $^{15}$N-acetonitrile as the hyperpolarization target. Using a $B_z = -15\ \mu T$, we find that a nearly four-fold enhancement over the SABRE-SHEATH signal may be obtained at an offset of $\pm 7\ Hz$ from the double quantum frequency for $^1$H-$^{15}$N at this field ($\omega_{DQ} = 574\ Hz$). By analyzing the population flow (ignoring exchange) in this system (Figure 4D), we find that two features stand out. Firstly, in contrast to the messy dynamics in Figure 3, the majority of the polarization comes from a difference between the $|S_H\alpha_L\rangle$ and $|S_H\beta_L\rangle$ state, which comes about as the $|S_H\alpha_L\rangle$ in the depicted case is close to an eigenstate of the propagator created by the circularly polarized experiment under these conditions while the $|S_H\beta\rangle$ state is heavily mixed with other states. Secondly, the sequence rearranges population out of the $|T_H^-\alpha_L\rangle$ state and into the $|T_H^0\alpha_L\rangle$ and $|T_H^+\alpha_L\rangle$ states, which are disconnected from the $|S_H\beta_L\rangle$ that serves as the source of hyperpolarization under these conditions. This circularly polarized SABRE experiment serves as a foundation for utilizing multi-axial magnetic fields to generate improved SABRE hyperpolarization. To further push the boundaries of what can be accomplished in this realm, we look to pulse shaping to obtain the control over the system dynamics required to further improve performance. It should be noted that this experiment is similar to current work where SABRE hyperpolarization is created using a $B_z = 50\ \mu T$ and $B_x$ field near the $^{15}$N-resonance, which is analogous to a LIGHT-SABRE experiment(25). However, that experiment utilizes $B_x \approx 1\ \mu T$, which simulates the conditions at high magnetic fields. This experiment, by contrast, uses two fields of identical magnitudes.

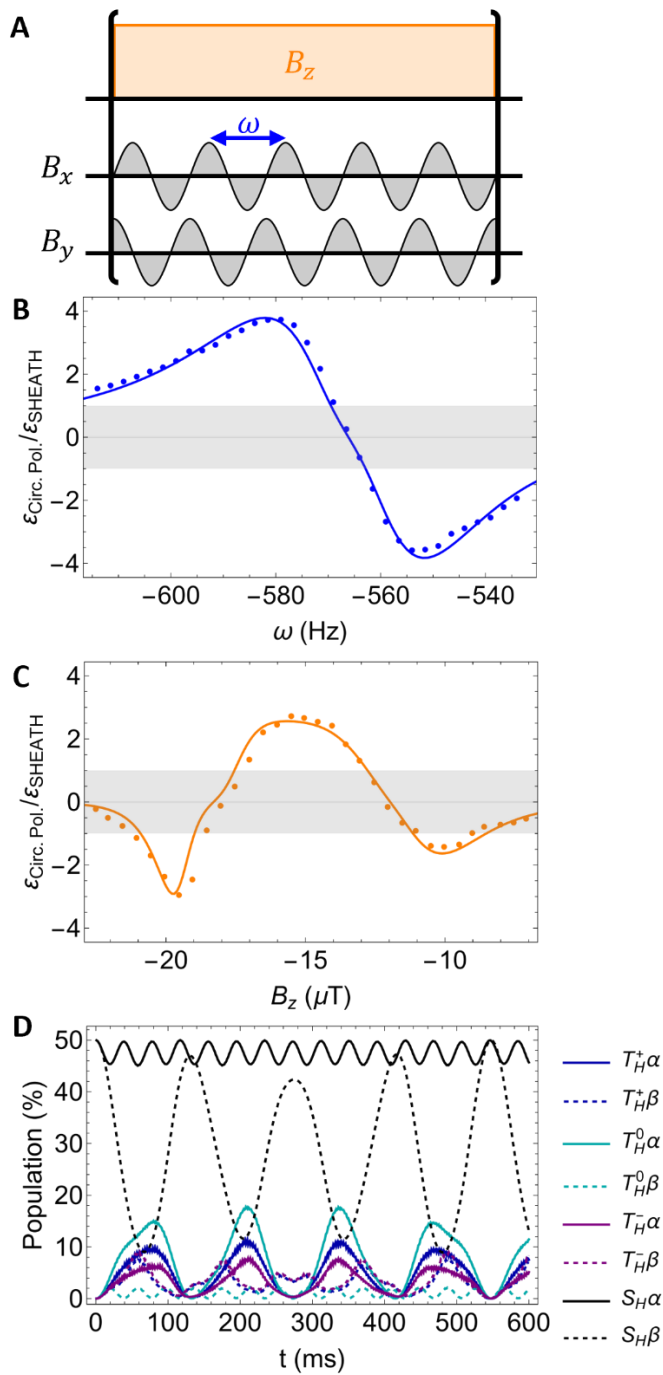

**Figure 4.** The circularly polarized SABRE experiment **A.** along with experimental validation of the improvement obtained relative to SABRE-SHEATH by **B.** sweeping the carrier frequency $\omega$ and **C.** leading field $B_z$. All experiments performed on $^{15}$N-acetonitrile. Data points are experimental, lines are simulation. **D.** This experiment improves hyperpolarization performance by selectively depopulating one of the two singlet states while preserving the other state in addition to rearranging the triplet populations such that exchange events cannot cause depolarization and drive the system backwards.

Pulse shaping affords flexibility to control complex systems in ways that may not be so simple to intuit. The traditional applications for pulse shaping in magnetic resonance mostly involve the development of single pulses that each accomplish a particular objective in controlling the system. For example, performing population inversion over a finite bandwidth and generating broadband excitation of magnetization with an identical phase are two canonical examples of pulse shaping in magnetic resonance [Pulse shaping citation]. SABRE offers a unique application for pulse shaping because its reversible nature allows the production of polarization to happen continuously over a minute as parahydrogen and target nuclei exchange with the catalyst, meaning that the shaped magnetic fields must be applied over this time. However, each "polarization transfer complex" (the active form of the SABRE complex where hyperpolarization occurs) is only present transiently in solution on timescales of tens of milliseconds, meaning that the shaped pulses must control the spin dynamics within millisecond timescales. In the following sections, we will describe the construction of the pulse shaping basis and optimization algorithm used in this work.

**Pulse shaping procedure and optimization** Periodic pulse shapes are ideal candidates to satisfy the unique timescale requirements introduced by SABRE as discussed above. We developed a custom Fourier basis set to shape each of the Cartesian field components, where we minimized the basis set dimensionality by directly shaping the relative amplitude and phase of the frequency components (see SI). Each waveform is a linear combination of an exponential decay in the frequency domain and a gaussian centered at an arbitrary frequency within the available bandwidth. The phase of each component was then toggled in a binary fashion to produce a wide variety of pulse shapes, which were then combined on all three field channels. To avoid inductive distortions introduced by the coil sets, the maximum frequency component of the waveform was restricted to 10 kHz. In addition to the pulse shape of each channel, the pulse amplitude as well as a constant magnetic field offset were allowed to vary with the shape. For the three magnetic field directions, this resulted in a total of 27 optimizable parameters for the simulation.

After establishing the basis for our pulse shapes, we utilized our SABRE model(24) as the basis of our optimization method. We chose to demonstrate this using $^{15}$N-SABRE and used a model system based on the magnetic and kinetic parameters of $^{15}$N-acetonitrile as a basis for this optimization (see SI). An evolutionary strategy (ES) optimization algorithm(26) was constructed to perform the optimization, as these methods avoid optimizing to local extrema and are particularly convenient for high-dimensional optimization problems. The ES algorithm that we used initializes by choosing 30 randomly sampled points in the pulse shape parameter space to be used as "parent" simulations. For each parent, the SABRE polarization is calculated after applying the pulse shape for 20 seconds and used to rank the parents, where only the top 5 parents are used as starting points for the subsequent generation and each parent generates an additional 30 children. This process is allowed to continue until either an optimum is reached, or a preset number of generations has passed. The pulse shape period was constrained to $T = 2.5\ ms$.

The optimized pulse shape is shown in Figure 5A. The ES algorithm produced the Multi-Axis Computer-aided HEteronuclear Transfer Enhancement (MACHETE) for SABRE pulse sequence, which utilizes only the $B_x$ and $B_z$ component of the magnetic field. This sequence generates a theoretical upper limit that improves the hyperpolarized signal by an order of magnitude. It is not intuitive that a large transverse magnetic field would be optimal for generating spin polarization as one would expect that the transverse field would readily destroy the spin polarization. However, this particular pulse shape preserves the z-projection of the $^{15}$N spin but rapidly nutates the $^{1}$H spin (Fig 5B). This feature of the pulse sequence is required so that the hyperpolarized signal is preserved and allowed to accumulate throughout the entire experiment.

In general, we will discuss the MACHETE shape in terms of the amplitude of the shaped field component, $A_n$, as well as the constant offset of that field component, $B_n$, as these are readily accessible experimental parameters. The optimal polarization was obtained with the parameters $A_x = 23.9\ \mu T$, $B_x = 5.2\ \mu T$, $A_z = -24.7\ \mu T$, and $B_z = -14.1\ \mu T$, which are the parameters used to run the simulation in Fig. 2. While the ES algorithm generated the MACHETE shape shown in Fig. 5 with $A_z < 0$ (the maximum field value of the shaped component is negative), we found that significantly enhanced hyperpolarization could also be obtained when $A_z > 0$, although performed slightly worse than the ES optimized shape, with a predicted enhancement over SABRE-SHEATH of 9.4-fold, in comparison to the 10.6-fold theoretical increase obtained with the ES optimized shape ($A_z < 0$). This required slight re-optimization of the magnetic field amplitudes, and optimal polarization could be obtained for this case with $A_x = 24.1\ \mu T$, $B_x = 1.25\ \mu T$, $A_z = 28.5\ \mu T$, and $B_z = -16.0\ \mu T$. In the following section, we will examine how these optimized shapes generate hyperpolarization.

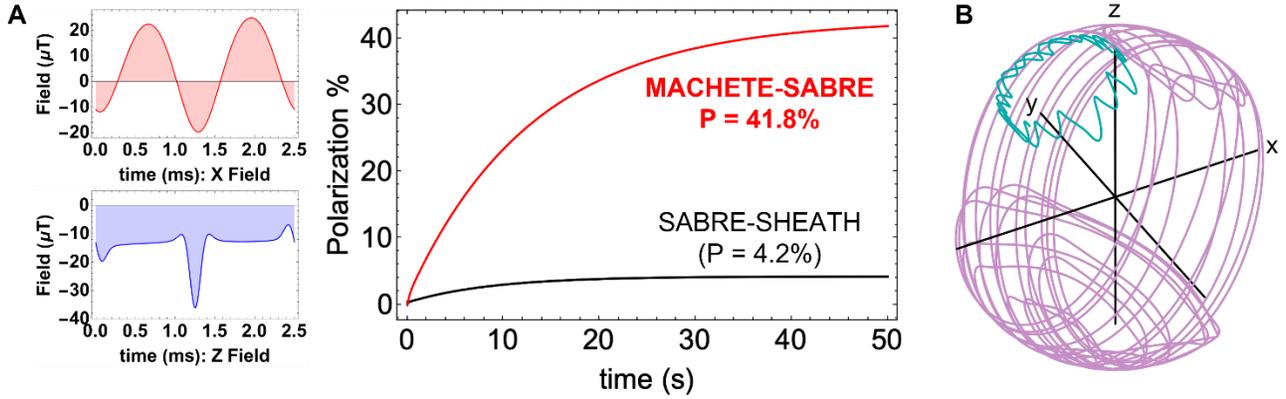

**Figure 5.** MACHETE-SABRE pulse sequence and experimental setup. **A.** The pulse shape optimized to use excitation with two shaped magnetic fields: a transverse magnetic field (that we denote $B_x$) and a longitudinal magnetic field, $B_z$, along which polarization is stored. Specific details as to the pulse parameters are detailed in the SI. After 50 seconds of simulations, the MACHETE-SABRE sequence generates an approximately 10-fold improvement over SABRE-SHEATH. **B.** Spin trajectories of $^{15}N$ (cyan) and $^{1}H$ (light purple) under 10 periods of the MACHETE-SABRE pulse sequence. While the z magnetization of the target nucleus is largely preserved for this sequence, which is required to preserve the generated once the target dissociates hyperpolarization, the proton magnetization is rapidly modulated.

**Analysis of the MACHETE-SABRE pulse shape** To understand the mechanism by which the MACHETE pulse sequence improves hyperpolarization performance, we tracked the flow of population in the system under the MACHETE pulse sequence and compared it to the flow of the same populations in the SABRE-SHEATH sequence (Fig. 6). While this procedure could be accomplished in any basis set, examining the flow of population in the STZ basis set (singlet-triplet on the parahydrogen derived hydrides, Zeeman basis on the target spin) provides physical relevance to the analysis, as the desired transition is between two states of this particular basis. In the SABRE-SHEATH experiment, both $|\alpha_L\rangle$ and $|\beta_L\rangle$ states are excited non-selectively, leading to complicated dynamics. In this case, net polarization is derived by the relatively small differences in the efficiency of pumping various transitions. In contrast, the ES-optimized MACHETE sequence selectively pumps the transition from $|S_H\alpha_L\rangle \rightarrow |T_H^+\beta_L\rangle$ while suppressing population flow to all other non-target states as well as preserving the $|S_H\alpha_L\rangle$ population. Similarly, the MACHETE shape with $A_z > 0$ generates a complete depletion of the $|S_H\beta_L\rangle$ and redistributes this population among only $|T_H^+\alpha_L\rangle$, $|T_H^0\alpha_L\rangle$, and $|T_H^-\alpha_L\rangle$, leading to a higher efficiency of generating hyperpolarization, as off-target excitations are suppressed. The state that is pumped by the J-couplings, $|T_H^-\alpha_L\rangle$, is actually populated the least so to minimize pumping the system back into the singlet state, whereas this feature is absent when $A_z < 0$. The slow-down in the population dynamics for MACHETE also allow for a greater fraction of the ligands that exchange to experience the generation of polarization in one direction, whereas under SABRE-SHEATH exchange at non-optimal times can invert the desired polarization. This would allow MACHETE to be more robust for different exchange rates, particularly when the lifetime is long, relative to SABRE-SHEATH.

In analyzing the performance of the MACHETE pulse sequence, we found that the timescale being considered is critical for understanding, predicting, and optimizing the performance of a SABRE sequence (Fig 7). While SABRE-SHEATH generates polarization much more rapidly than the MACHETE sequence, the original experiment readily destroyed magnetization if the system was over-driven. To correct that, we previously introduced the coherently-pumped SABRE-SHEATH experiment(27), which interleaves period of evolution to generate hyperpolarization with delay periods to allow for exchange. This strategy mitigated back-pumping of polarization by interleaving periods to allow for substrate to exchange, which generated 2- to 3-fold improvements over the traditional SABRE-SHEATH sequence. Interestingly, the MACHETE sequence does not perform as well compared to either of these sequences for times on the order of a few exchange events. Despite this, the MACHETE sequence outperforms the SABRE-SHEATH experiments on timescales far longer than both the coherent dynamics as well as exchange, which is likely permitted due to the high preservation of singlet order in the MACHETE sequence. This provides new insight into optimization methods of other areas of SABRE experiments.

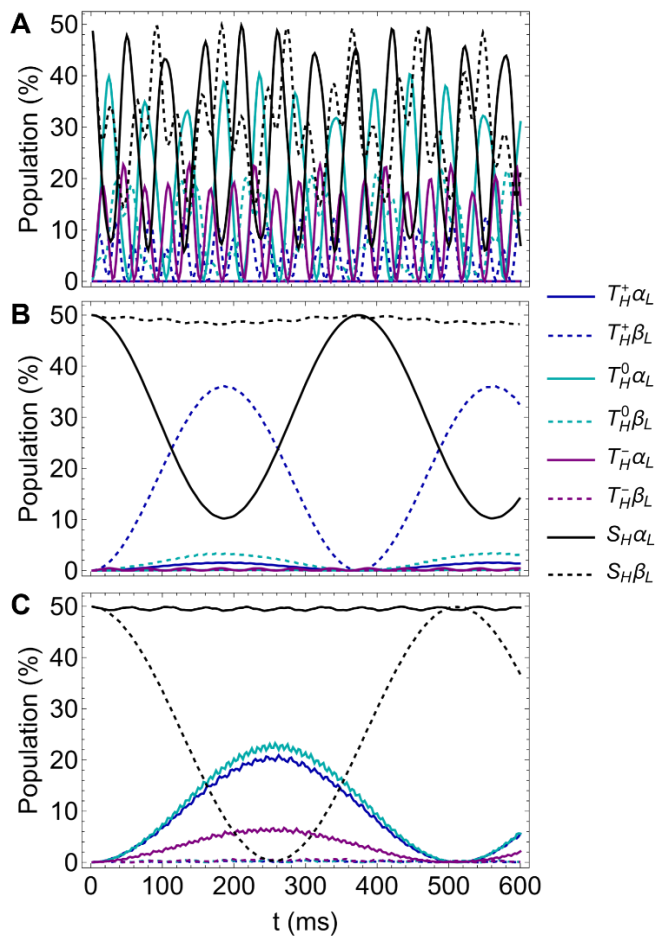

**Fig. 6.** Population flow in **A.** SABRE-SHEATH, **B.** the ES optimized MACHETE shape with $A_z < 0$, and **C.** the MACHETE shape with $A_z > 0$. In SABRE-SHEATH, population flows out of both singlet states into both target state as well as off-target states, making it a relatively inefficient process. In contrast, the ES optimized MACHETE shape selectively generates population transfer between a single pair of states ($|S_H \alpha_L\rangle \rightarrow |T_H^+ \beta_L\rangle$) while still preserving population in the $|S_H \beta_L\rangle$ state. Similarly, applying the MACHETE shape with $A_z > 0$ completely depletes the $|S_H \beta_L\rangle$ state and distributes population into $|T_H^+ \alpha_L\rangle$, $|T_H^0 \alpha_L\rangle$, and $|T_H^- \alpha_L\rangle$. Exchange is suppressed in these calculations.

**Experiments with the MACHETE sequence** These theoretical conclusions were experimentally validated by applying the MACHETE sequence to two model systems: $^{15}$N-acetonitrile and $^{15}$N-pyridine. For both systems, the optimal field conditions for SABRE-SHEATH were determined by sweeping a static average $B_z$ field. Then, the MACHETE field offsets $B_x$ and $B_z$ were swept while shaped field amplitudes were held constant to determine the optimal field conditions for MACHETE. Parameters for sample preparation and data collection may be found in SI.

For $^{15}$N-acetonitrile, the $A_z < 0$ MACHETE shape generated a 42655-fold enhancement over thermal signal at 1T

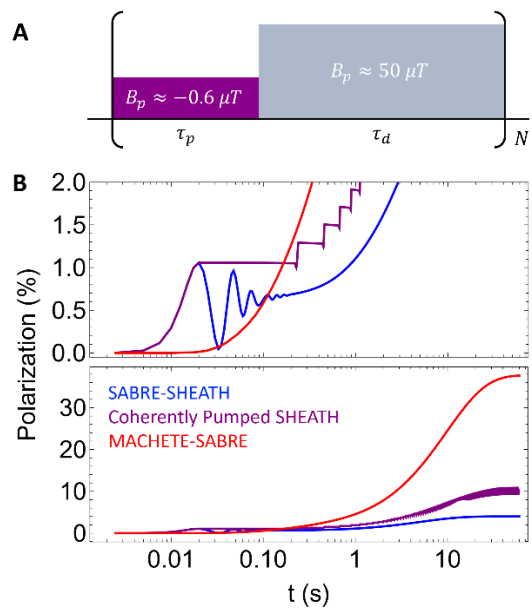

**Fig. 7.** The performance of SABRE experiments varies depending on timescale, where mediation of back-pumping was first accomplished with **A.** the coherently-pumped SABRE-SHEATH experiment. **B.** Compared to SABRE-SHEATH, the coherently pumped SABRE-SHEATH variant prevents losses from over-driving the system and losing polarization. Despite not generating as much polarization as either SABRE-SHEATH or the coherently pumped variant over timescales of a few exchange events, the MACHETE-SABRE sequence outperforms any other sequence substantially at long times.

at the conditions $B_x = 4.25\mu T$ and $B_z = -14.10\mu T$, representing a 9.7-fold improvement over optimal SABRE-SHEATH measurements taken on the same experimental day (4400-fold enhancement at $B_z = -0.52\mu T$). The experimental improvement in hyperpolarized signal validates the approximate ten-fold improvement over SABRE-SHEATH observed in simulation. Furthermore, the MACHETE-derived polarization can generate more polarization than SABRE-SHEATH over ~2-3 μT ranges about the optimal conditions (Fig 8), indicating that the MACHETE sequence is robust to field inhomogeneities and variations.

Despite not having optimized the MACHETE sequence to systems such as $^{15}$N-pyridine that binds two targets simultaneously to the equatorial plane of the iridium complex, we were still able to demonstrate significant improvements to $^{15}$N-pyridine hyperpolarization with MACHETE-SABRE. In $^{15}$N-pyridine, $A_z < 0$ MACHETE produced a 16026-fold enhancement over thermal at $B_x = 4.25\mu T$ and $B_z = -14.10\mu T$, which corresponds to a 6.1-fold improvement over optimal SABRE-SHEATH ($\varepsilon = 2623, B_z = -0.52\mu T$).

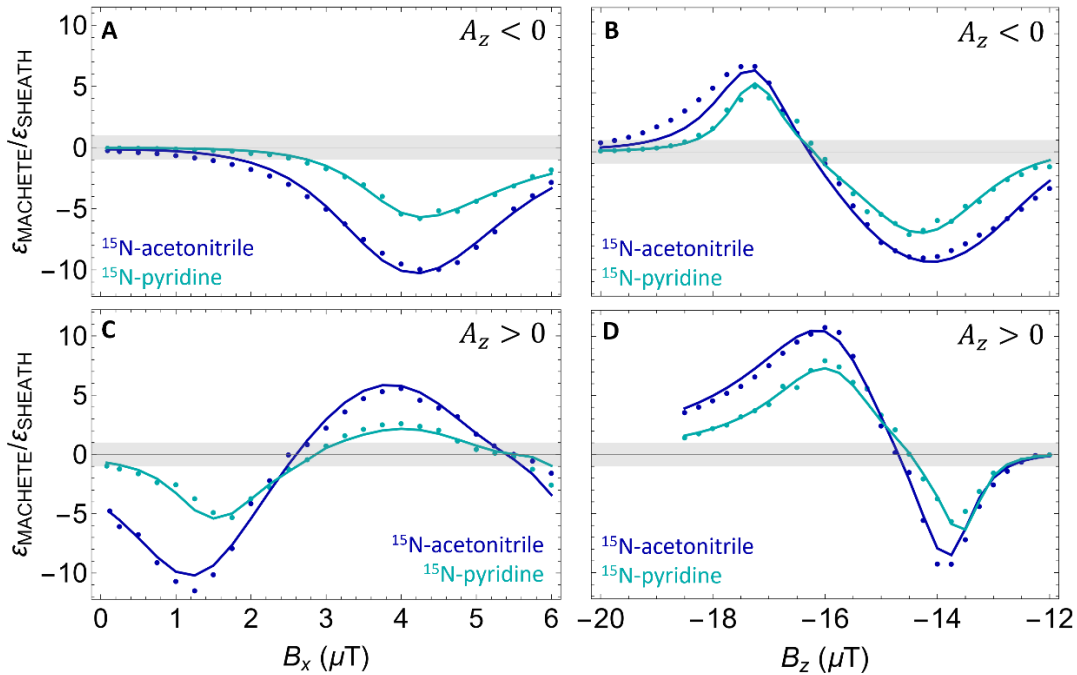

**Fig. 8.** MACHETE-SABRE signal enhancement over SABRE-SHEATH for both $^{15}$N-acetonitrile and $^{15}$N-pyridine at various field conditions. For the $A_z < 0$ shape, $B_x$ sweeps **(A)** were conducted with constant $B_z = -14.10\mu T$ while $B_z$ sweeps **(B)** were conducted with constant $B_x = 4.25\mu T$ for both acetonitrile and pyridine. The $A_z > 0$ sweeps were conducted at different fields for both model systems. For the $A_z > 0$ $B_x$ sweeps **(C)**, $^{15}$N-acetonitrile held a constant $B_z = -13.70\mu T$ while 15N-pyridine used $B_z = -16.00\mu T$. Similarly, for the $B_z$ sweeps **(D)**, $^{15}$N-acetonitrile used a constant $B_x = 1.25\mu T$ while 15N-pyridine held $B_x = 2.00\mu T$ MACHETE-SABRE hyperpolarized signal (blue for $^{15}$N-acetonitrile and cyan for $^{15}$N-pyridine) demonstrated improvement over optimal SABRE-SHEATH (gray) at numerous field conditions. Experimental data is represented by points while simulation data is shown as solid lines.

Similar magnitudes of hyperpolarization were observed with the $A_z > 0$ MACHETE shape for both $^{15}$N-acetonitrile and $^{15}$N-pyridine. $A_z > 0$ MACHETE produced an enhancement over thermal of 34743 and 22791 for $^{15}$N-acetonitrile and $^{15}$N-pyridine, respectively. This corresponds to a 9.6-fold improvement over SABRE-SHEATH ($\varepsilon = 3514$) for $^{15}$N-acetonitrile and a 7.2-fold improvement over SABRE-SHEATH ($\varepsilon = 3156$) for $^{15}$N-pyridine. Interestingly, the ES optimized sign convention performed worse for the $^{15}$N-pyridine system than the $A_z > 0$ shape. This may be due to the suppression of $|T_H^- \alpha_L\rangle$ seen in Fig 6C since this state would shuttle polarization away from the target nucleus in a more highly coupled system.

Importantly, the MACHETE sequence was optimized on an ideal, model system without ancillary $^1$H nuclei coupled into the target nucleus to disrupt or attenuate polarization flow. When translating this experiment to real systems, the improvements predicted from the ideal case closely reflect the actual experimental data (Fig. 8). The simulations we show are using the small, model systems and show both nearly identical behavior and relative performance to what we observed in experiments. This means that more computationally expensive simulations optimizing pulse shapes to specific targets are not necessary to generate large improvements in the hyperpolarization. It is important to note that experimental yield is lower than in the simulations shown in Fig 2A because hydrogen exchange is a limiting factor. Despite this, finding ways to increase the parahydrogen association rate would provide a route to permit SABRE hyperpolarization on the order of conventional dDNP experiments.

**Conclusions**

Here, we have demonstrated that utilizing ultralow, shaped magnetic fields to design, optimize, and perform SABRE experiments can improve polarization by an order of magnitude. We developed the MACHETE-SABRE sequence from numerical optimization of SABRE dynamics, and showed that the improvement in performance arises both from the selectivity of the transition being excited in the SABRE system as well as preventing depletion of the parahydrogen order throughout the experiment. These results were confirmed experimentally on both $^{15}$N-acetonitrile and $^{15}$N-pyridine, which both showed significant improvements under the MACHETE sequence relative to SABRE-SHEATH. Furthermore, improved polarization could be generated over few-microTesla bandwidths, emphasizing the robustness of the MACHETE pulse sequence with respect to experimental imperfections.

## Conflicts of interest

There are no conflicts to declare.

## Acknowledgements

This work was supported by the National Science Foundation grant CHE-2003109.